# Multiple rooks of chess - a generic integral field unit deployment technique

Sabyasachi Chattopadhyay[a,*], A. N. Ramaprakash[a], Pravin Khodade[a], Kabir Chakrabarty[a], Shabbir Shaikh[a], Haeun Chung[b,c], and Sungwook E. Hong[d]

[a]Inter-University Centre for Astronomy and Astrophysics, Pune, India
[b]Korea Institute for Advanced Study, 85 Heogiro, Dongdaemun-gu, Seoul 02455, Korea
[c]Department of Physics and Astronomy, Seoul National University, 1 Gwanak-ro, Gwanak-gu, Seoul 08826, Korea
[d]Korea Astronomy and Space Science Institute, 776 Daedeokdae-ro, Yuseong-gu, Daejeon 34055, Korea

**Abstract**

A new field re-configuration technique, Multiple Rooks of Chess, for multiple deployable Integral Field Spectrographs has been developed. The method involves a mechanical geometry as well as an optimized deployment algorithm. The geometry is found to be simple for mechanical implementation. The algorithm initially assigns the IFUs to the target objects and then devises the movement sequence based on the current and the desired IFU positions. The reconfiguration time using the suitable actuators which runs at 20 cm/s is found to be a maximum of 25 seconds for the circular DOTIFS focal plane (180 mm diameter). It is similar to some of the fastest schemes currently available. The Geometry Algorithm Combination (GAC) has been tested on several million mock target configurations with object-to-IFU ($\tau$) ratio varying from 0.25 to 16. The configuration had both contiguous and sparse distribution of targets. The MRC method is found to be extremely efficient in target acquisition in terms of field revisit and deployment time without any collision or entanglement of the fiber bundles. The efficiency of the technique does not get affected by the increase of number density of target objects. For field with $\tau > 1$ prioritization of target objects is an optional feature and not necessary. The GAC can be modified for an instrument with higher or lower number of IFUs and different field size without any significant change in the flow. The technique is compared with other available methods based on sky coverage, flexibility and overhead time. The proposed geometry and algorithm combination is found to have advantage in all of the aspects.

## 1   INTRODUCTION TO TARGET ACQUISITION ALGORITHMS

In a multi-object fiber based spectrograph, acquisition of field targets is achieved by movement of the fibers and/or fiber bundles. Movement of multiple IFUs on the focal plane would require an intelligent algorithm to work without human intervention for efficient field reconfiguration. A primary need of any Target Acquisition Algorithm (TAA) is to minimize the time required to acquire the targets within the mechanical deployment constraints imposed by the instrument. Often the target field reconfiguration is required to be achieved within the detector readout time and/or within the telescope slewing time.

Different algorithms exist for movement of IFUs, which can be broadly classified into two categories: sequential and parallel [1]. The Two Degree Field survey (2dF [2]) at Anglo Australian Observatory, Automatic

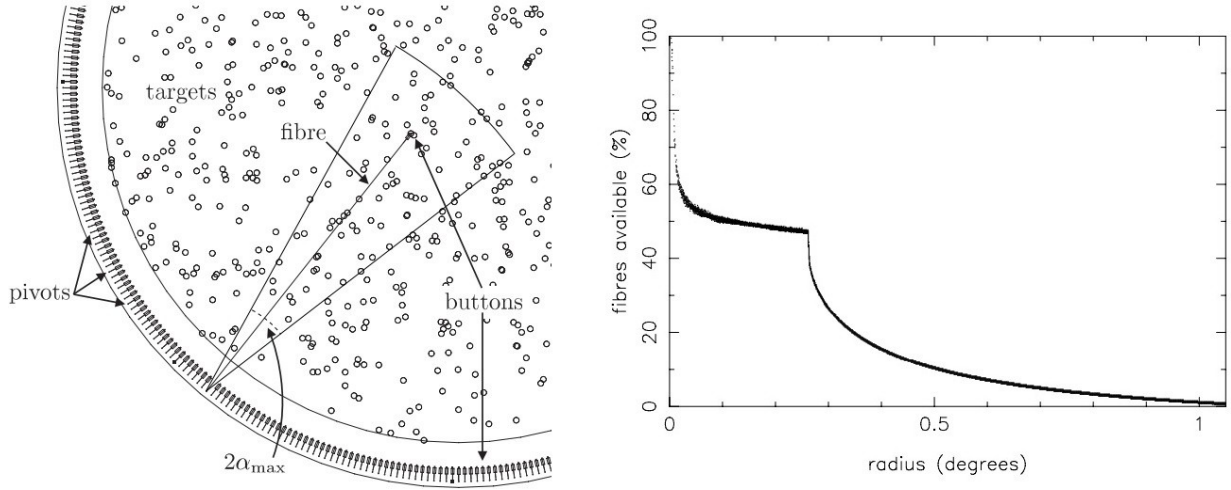

Figure 1: The left panel shows a 2dF field plate and different components are marked for field reconfiguration. Each fibre placement is constrained within the outlined sector that subtends an angle $2\alpha_{max}$. In the right panel, the target accessibility percentage of fibres against the radial distance from the field centre for 2dF is shown (empirically derived), which shows a bias towards central targets. The images are taken from [7].

Fiber Positioner (AUTOFIB [3]) at William Herschel Telescope, Hydra at Kitt Peak National Observatory[4] uses sequential movement with the fishermen around pond geometry. In this geometry, the fibers (fishermen) sit around the FOV (pond). The fibers can be drawn from the edge to any position on the FOV. On the other hand, the FMOS Echidna at the Subaru Telescope [5] and the Starbugs used for the TAIPAN instrument at the UK Schmidt telescope [6] employ parallel movement by densely packing the individual fiber deployment units. Compared to the sequential techniques, the parallel techniques are much faster in deployment at the cost of contiguous sky coverage limitations imposed by their geometry.

## 1.1 Sequential Technique - Fishermen Around Pond

The fishermen around pond geometry is used to deploy individual fibers over a field, where the fibers can crossover each other, as the out of plane movements of the fiber input tips are allowed. This geometry allows for the movement of the fibers within a particular angle (sector angle - $2\alpha$). As a result, a larger number of targets can be allocated closer to the center, when compared to farther towards the edges as shown in Figure 1. This signifies a bias towards centrally clustered objects. Prioritization of the targets is required to counter this bias. The number of target objects in a field ($N_T$) can be different from the number of sampling elements - SE ($N_{SE}$), e.g., the fibers or the IFUs. Their ratio $\tau$ ($= N_T/N_{SE}$) can be a significant number, if there are priorities assigned to any object (as in the case of 2dF).

Simulated Annealing (SA) describes a way out in such constraints. It is primarily a random walk in the parameter space $S$ to find the solution in an 'exhaustive' manner guided by a well-defined objective function $f$ without any insight of the target distribution. Here, $f$ is the function to be optimized, out of which $S$ is the space of all possible solutions. SA is resource intensive because it is exhaustive in nature, which may be critical for flexibility of observation. Due to its sequential nature, SA takes ~1 hour to reconfigure a 2dF field consisting of 400 fibers.

One of the major limitations of this method is the need to perform collision detection before performing the fiber deployment. SA pre-calculates all the possible collisions for a target field and stores them in an indexed manner which is called the collision matrix [7]. Based on fiber reach, fiber proximity, etc. the SA algorithm then goes on to check whether the allocation is valid or not. The validation process is extremely time-consuming. It would be a much better option to use a scheme that avoids collision inherently, compared to the option of going through all possible solutions and checking each solution against the collision matrix.

## 1.2 Parallel Techniques

The techniques which use Starbugs, Echidna or bend arm robots can move all or at least several fibers together. Starbugs uses multiple piezoceramic concentric tubes to hold and position of the fibers on the focal plane. Echidna uses tilted spine technology where the robots are fixed but the fiber tip can reach out to the objects within the patrol area. Multi Object Optical & Near infrared Spectrograph (MOONS) on Very Large Telescope (VLT) [8] uses bend arm technology. In case of MOONS, the fibers are held by tubes and routed along z axis out of the field. The bend arm robots responsible for movements, occupy a space above the focal plane. In Echidna and bend arm techniques, each fiber movement is restricted to a small part of the entire patrol field (1.2% for Echidna and 0.2% for MOONS). Several hundreds of fibers and robotic arms are densely packed to cover the entire area. The Starbugs robots can travel to a slightly larger part of the patrol field ( 9% for TAIPAN, 2.5% for MANy-Instrument FibEr-positioning SysTem - MANIFEST). The parallel methods are very fast compared to sequential techniques. The drawback of these schemes is their inability to acquire targets which are clustered individual targets in the sky. This limitation exists due to the geometry of densely packed robots, whose diameter is at least an order of magnitude higher than the fiber diameter. Similar to the sequential techniques, the parallel schemes require prioritization of target objects to avoid their limitations.

A new geometry and an algorithm is developed to overcome the current limitations of the field reconfiguration techniques. In this chapter, we will describe our Multiple Rooks of Chess algorithm and its generic nature (section 2). In section 3, we will show different simulation results to analyze the efficiency and compare them with other algorithms.

## 2 MULTIPLE ROOKS OF CHESS

### 2.1 Motivation

Since the geometry is a part of the observation strategy, an algorithm can only optimize what the geometry provides. Rather than a discussion of an algorithm only, it is apt to discuss the pros and cons of the geometry-algorithm combination (GAC). Any GAC should take care of efficiency, uniformity of target acquisition and feasibility of observation. Need for artificial prioritization of targets should be avoided, and the GAC should be able to acquire any target distribution. Since each GAC has to work within a constrained structure, the achievable optimization is also constrained.

The GAC presented here has been built for the Devasthal Optical Telescope Integral Field Spectrograph (DOTIFS) [9] to be used on the 3.6m Devasthal Optical Telescope (DOT) [10]. DOTIFS deploys 16 Integral Field Units (IFU) in a 4x4 grid, over a field of view (FOV) of 8' diameter (180mm physical diameter). Any number of IFUs can form a conglomerate on the focal plane or can be distributed in the field with a minimal gap of 0.5" between them. This provides for continuous coverage of an extended object as well as for up to sixteen discrete regions of interest within the field of view. The GAC proposed here can be extended to a larger or a smaller grid with varying sizes and shapes of the IFUs and the focal plane.

### 2.2 Geometry

We propose a GAC which is entirely different from the present day techniques of target acquisition. The out of plane movement of the sampling elements (SE) only increases the possibility of collision and fiber entanglement. On the other hand, in-plane movement systems would enforce further constraints on the movement of the IFUs, which can be compensated by an efficient allocation and movement strategy. The method of radial movement is simple to implement but it introduces a bias towards the central targets. Hence, the movement of any SE has been decided to be along either the y or the z direction. Thus, the sampling elements would travel either in the y or the z direction, similar to the movement of the rooks in chess and hence the name, Multiple Rooks of Chess (MRC). We have used orthogonally mounted linear stages on both sides and at different heights from the focal plane as shown in Figure 2. The z positioners do not travel into the focal plane, but carry the y linear stages which hold the SEs on their tip as shown in Figure 3. The number of the y and z linear stage assemblies (here after orthogonal assemblies - OA) on each layer of each side (which is called a Group, shown as A, B, C & D in Figure 2) is important as any out of plane movement of the SEs are not allowed. The OAs and the associated

SEs that belong to a Group cannot alter their relative z positions, but their positions are independent along y direction. The number of Groups are also important as OAs that belong to different Groups cannot crossover to acquire objects having the same z position. Within this geometry, the number of Groups and the number of OAs in a Group is derived from the science requirements. The algorithm part is able to allocate the SEs to the targets for any combination of the number of Groups and the number of OAs in a Group. The groups and their associated SEs are shown in Figure 4.

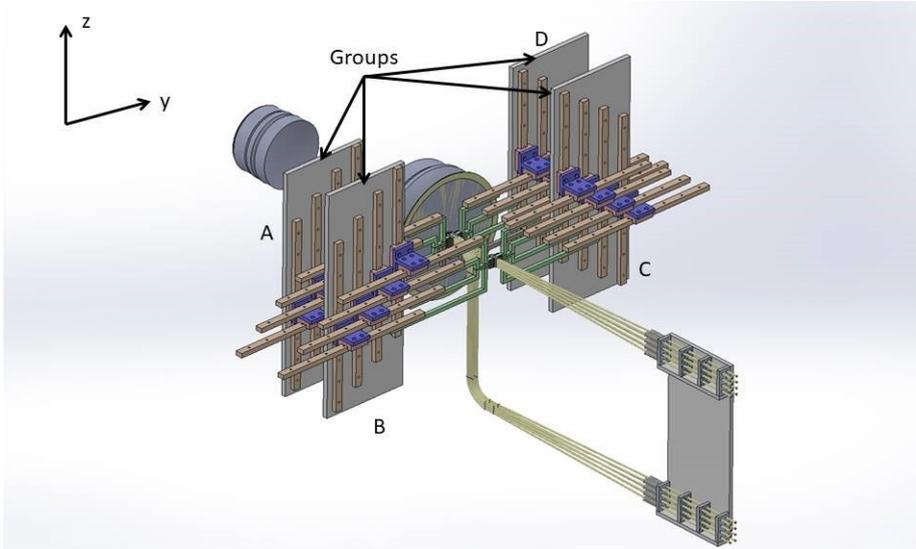

Figure 2: Figure shows an illustration of the distribution of the linear stages around the focal plane the definition of Groups. Four Groups each containing four y and z positioner assemblies are mounted on two sides of the focal plane at a different height from the focal plane. More Groups can be placed in the architecture by adding more layers on each side. A, B, C, D represents the mounting of the four groups each having four y z linear stage assemblies.

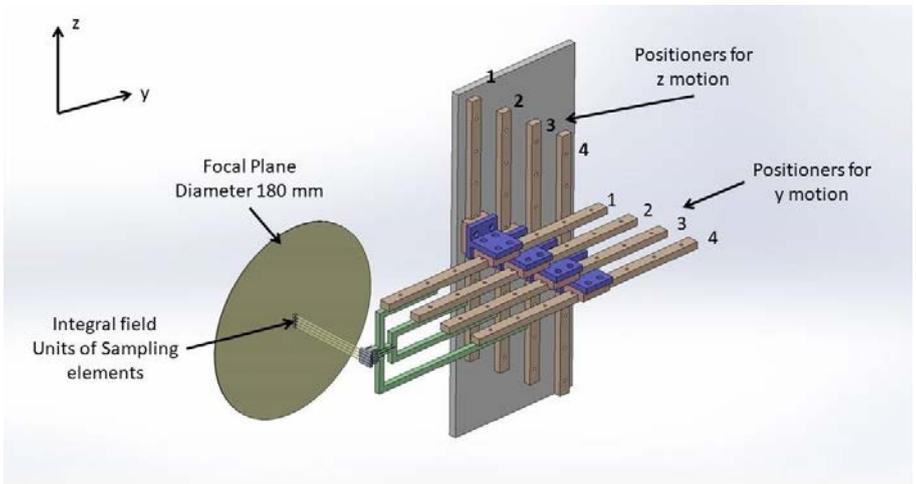

Figure 3: Figure shows an illustration of the y z assembly. A Group consisting of 4 y z positioner assemblies holding 4 sampling elements on the focal plane. The number of OAs per Group can be flexible depending on the mechanical constraints of the design. 1, 2, 3, and 4 are the four assemblies in order of decreasing $y$ coordinates

Figure 4: Naming convention of sampling elements.

## 2.3 Algorithm

The algorithm is primarily divided into three parts. In the first part, the SEs are assigned to the Target Objects based on the mechanical constraints from the Geometry. Once they are assigned, movement of the SEs is devised based on their current (initial) configuration and target (final) configuration. In the final step, the SEs are moved with the help of their respective OAs.

During the assignment procedure, allocation of the Targets to the SEs is performed for one Group at a time. The objects are assigned to the lowest y group (group A) first and subsequently to other groups with higher y coordinate. For assigning the targets to any group, a set of objects $S^T$, are pooled in as available targets from the mother set $S^A$ of all available targets. The process of pooling is done in the following way. From the $S^A$, lowest y objects are added to the $S^T$ until the number of objects in the $S^T$ becomes equal to the number of OAs in the Group. It is ensured that no two objects in the $S^T$ have a z overlap to comply with the mechanical constraints. Once the number requirement is fulfilled, the object coordinates are sorted in the order of their z coordinates and assigned to the SEs of the Group. The object with highest z coordinate gets assigned to the SE with highest z coordinate and so on. This procedure is then sequentially extended for all the groups (B, C, D in that order) until the $S^A$ gets exhausted. The highest number of objects which can share an overlapping z position is equal to the number of Groups. Two SEs are overlapping if the difference between the z coordinate of the central spaxel of them is lower than an SE width (3.2 mm). This leads to an inability to acquire a target configuration in some cases. The limitation is overcome by rotating the field relative to the deployment mechanism. An SE is uniquely named by combining the group number (A, B, C, D shown in Figures 3) with the OA order (1, 2, 3, 4 shown in 2) as shown in 4. An example of the assignment process is shown in Figure 5 and the corresponding steps are mentioned below.

- Step 1: The y and z coordinates are collected from user for all targets.
- Step 2: Target with lowest y coordinate from the available targets in $S^A$ is selected for group A. Total pooled targets for group A is now 1.
- Step 3: Target with lowest y coordinate from the available targets in $S^A$ is selected for group A. Total pooled targets for group A is now 2.

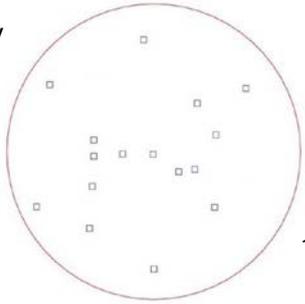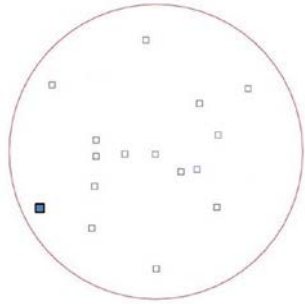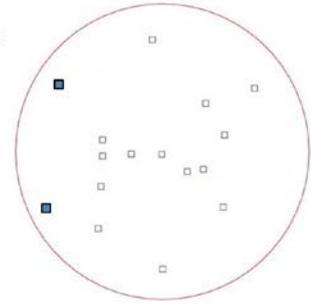
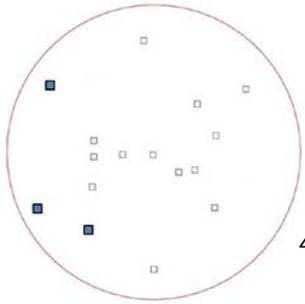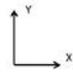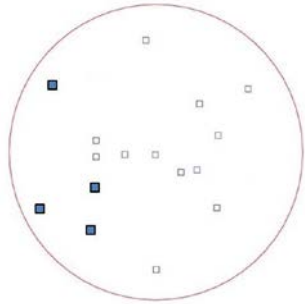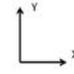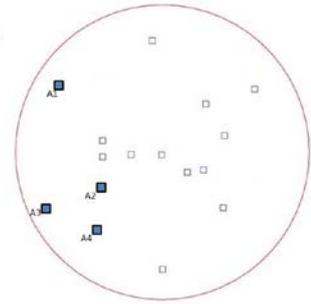
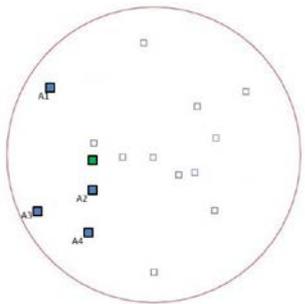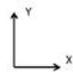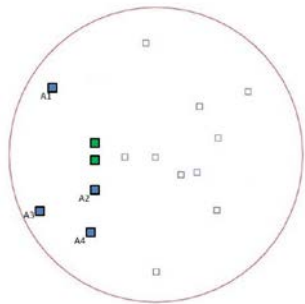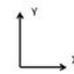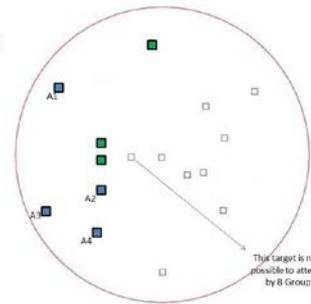
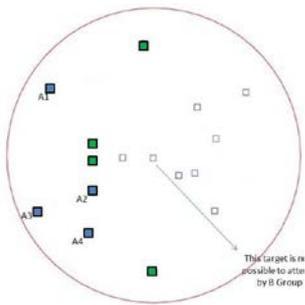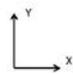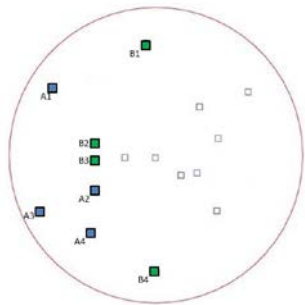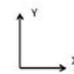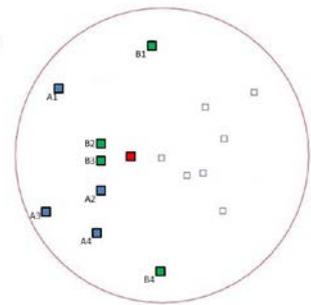
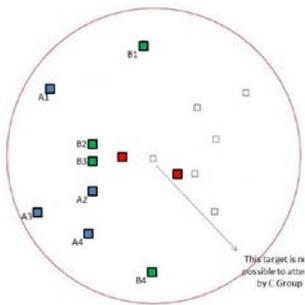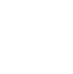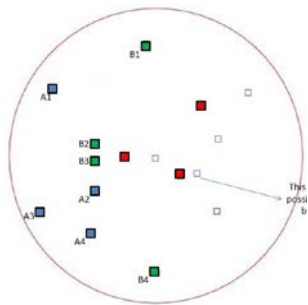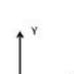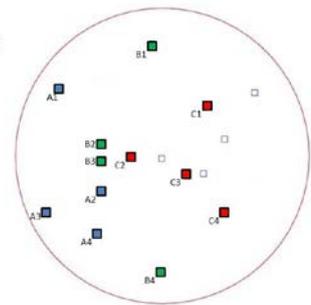

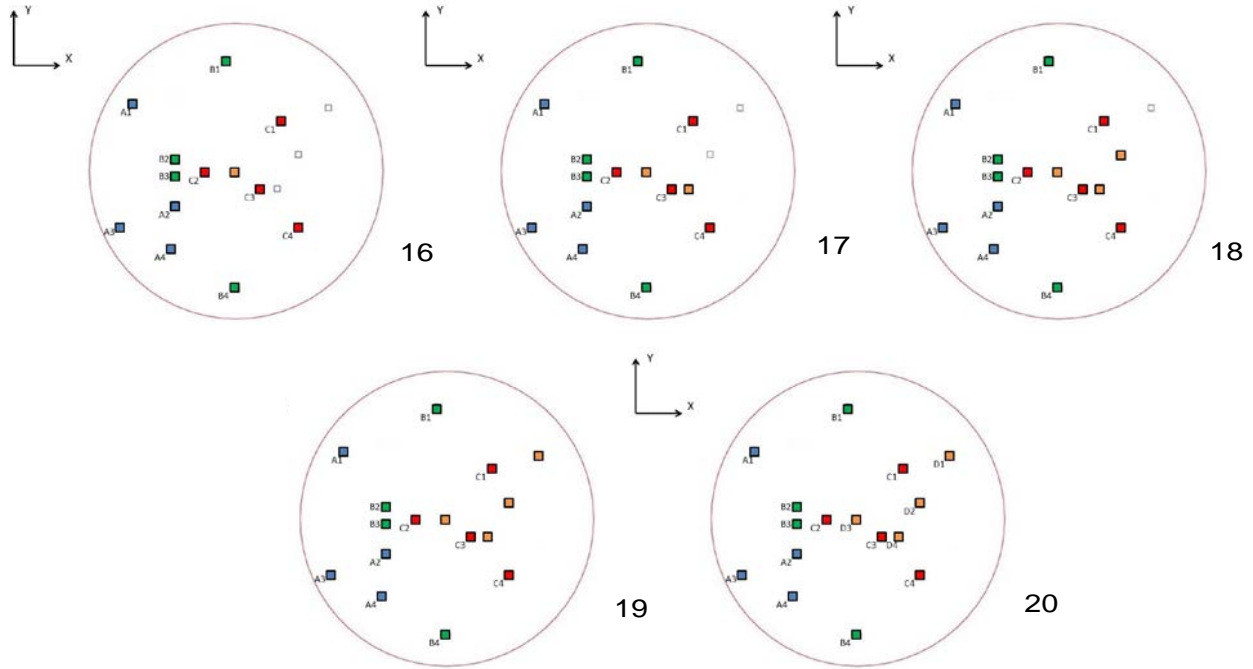

Figure 5: Steps for assigning of 16 sampling elements to 16 target objects (mock field) is illustrated. The circle represents the field of view and the empty rectangles represents the position of a target. The colored rectangle represents the SEs position after assignment. different color represents different groups of SE.

- Step 4: Target with lowest y coordinate from the available targets in $S^A$ is selected for group A. Total pooled targets for group A is now 3.

- Step 5: Target with lowest y coordinate from the available targets in $S^A$ is selected for group A. Total pooled targets for group A is now 4. No more targets will be selected for Group A since the number of targets selected for this group is now equal to the number of SEs in the group.

- Step 6: The 4 targets are sorted in the order of their z coordinate and allotted to SEs of group A. The object with highest z coordinate gets assigned to the SE with highest z coordinate and so on.

- Step 7: Target with lowest y coordinate from the available targets in $S^A$ is selected for group B. Total pooled targets for group B is now 1.

- Step 8: Target with lowest y coordinate from the available targets in $S^A$ is selected for group B. Total pooled targets for group B is now 2.

- Step 9: Target with lowest y coordinate from the available targets in $S^A$ cannot be selected for group B as it has z overlap with target selected in step 7. Target with second lowest y coordinate from the available targets is selected for group B. Total pooled targets for group B is now 3.

- Step 10: Target with lowest y coordinate from the available targets in $S^A$ cannot be selected for group B as it has z overlap with target selected in step 7. Target with second lowest y coordinate from the available targets is selected for group B. Total pooled targets for group B is now 4. No more targets will be selected for Group B since the number of targets selected for this group is now equal to the number of SEs in the group.

- Step 11: The 4 targets are sorted in terms of their z coordinate and allotted to SEs of group B. The object with highest z coordinate gets assigned to the SE with highest z coordinate and so on.

- Step 12: Target with lowest y coordinate from the available targets in $S^A$ is selected for group C. Total pooled targets for group C is now 1.

- Step 13: Target with lowest y coordinate from the available targets in $S^A$ cannot be selected for group C as it has z overlap with target selected in step 12. Target with second lowest y coordinate from the available targets is selected for group C. Total pooled targets for group C is now 2.

- Step 14: Target with lowest y coordinate from the available targets in $S^A$ cannot be selected for group C as it has z overlap with target selected in step 12. Target with second lowest y coordinate from the available targets is selected for group C. Total pooled targets for group C is now 3.

- Step 15: Target with lowest y coordinate from the available targets in $S^A$ is selected for group C. Total pooled target for group C is now 4. No more targets will be selected for Group C since the number of targets selected for this group is now equal to the number of SEs in the group. The 4 targets are sorted in terms of their z coordinate and allotted to SEs of group C. The object with highest z coordinate gets assigned to the SE with highest z coordinate and so on.

- Step 16: Target with lowest y coordinate from the available targets in $S^A$ is selected for group D. Total pooled targets for group D is now 1.

- Step 17: Target with lowest y coordinate from the available targets in $S^A$s is selected for group D. Total pooled targets for group D is now 2.

- Step 18: Target with lowest y coordinate from the available targets in $S^A$ is selected for group D. Total pooled targets for group D is now 3.

- Step 19: Target with lowest y coordinate from the available targets in $S^A$ is selected for group D. Total pooled targets for group D is now 4. No more targets will be selected for Group D since the number of targets selected for this group is now equal to the number of SEs in the group.

- Step 20: The 4 targets are sorted in terms of their z coordinate and allotted to SEs of group D. The object with highest z coordinate gets assigned to the SE with highest z coordinate and so on.

After completion of object assignment to SEs, the movement sequence is devised based on the available space for each SE to move. The available space for an SE is defined by the previous and the target configuration. The principle of the algorithm relies on avoiding a collision between any two SEs. An SE (and the relevant OA) can move around without a collision in the open spaces on the focal plane. Hence, maneuvering the movement of the SEs through the open spaces is the key to devise the movement sequence. At the first step, the algorithm defines the boundary for all the groups. During this procedure the SEs may be assigned to a temporary location at some distance from its position. This process is executed for both initial and final configuration. Hence, apart from the initial and final position, the algorithm also generates temporary initial and temporary final position. The algorithm first moves the SEs from initial to temporary initial position. A temporary initial position is shown in Figure 7. Movement from temporary initial position to temporary final position is defined by the space an SE has for movement. This sequence generation process is shown later. After this the SEs move from temporary final position to final position. A final position is shown in step 20 of Figure . Through this process, the algorithm creates a temporary initial configuration and a temporary final configuration based on temporary initial positions and temporary final positions respectively.

The process of making a temporary initial configuration or a temporary final configuration is described below. The process illustrated in Figure 6 is for creating the temporary final configuration.

- Step 1: For group A, the y coordinate of the highest y SE is defined as the boundary. The boundary is marked with a vertical blue line.

- Step 2: Group B SEs B2 and B3 are moved to a temporary position outside the boundary of group A.

- Step 3: For group B, the y coordinate of the highest y SE is defined as the boundary. The boundary is marked with a vertical green line.

- Step 4: Group C SE C2 is moved to a temporary position outside the boundary of group B.

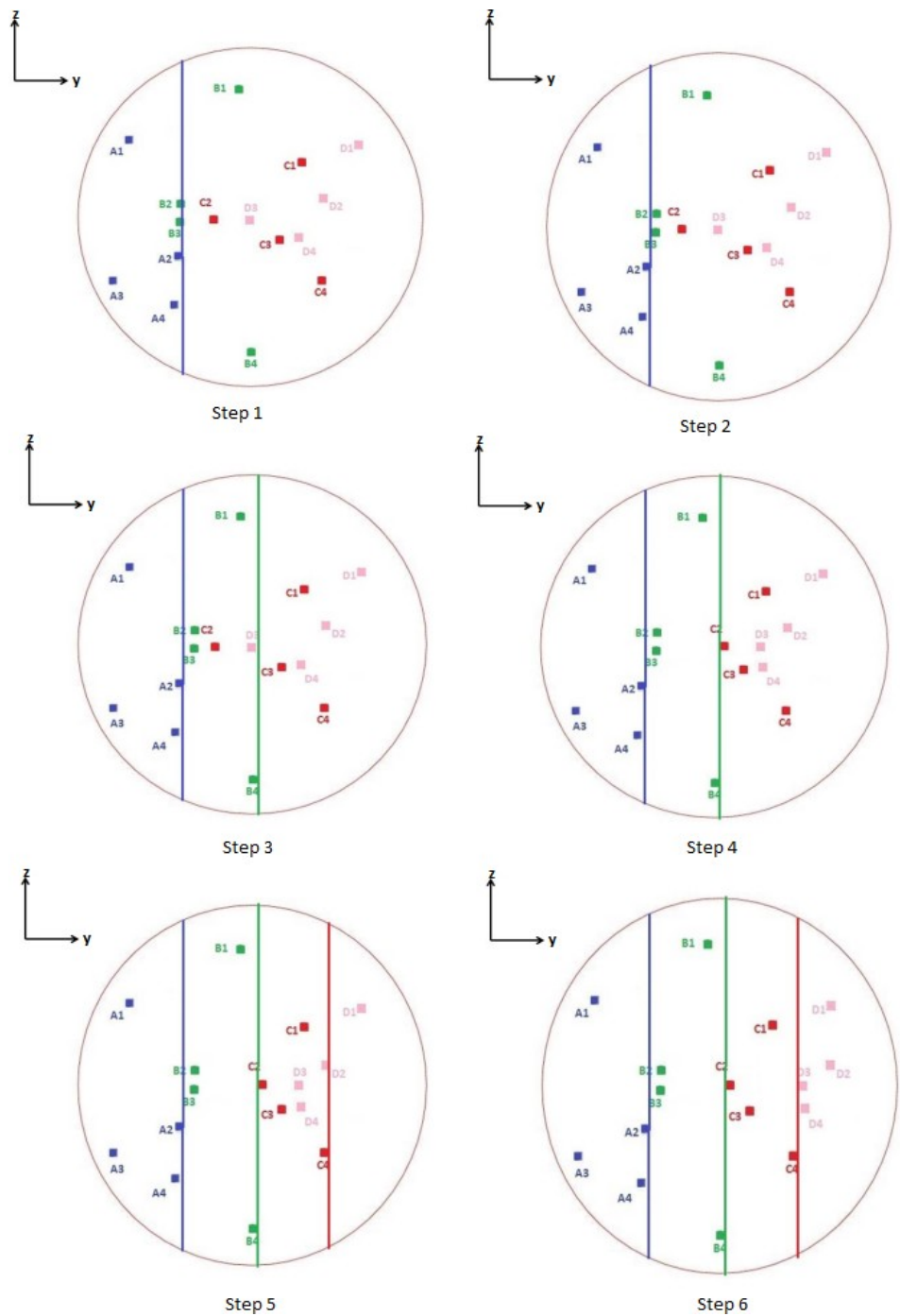

Figure 6: Illustrates the sequence of of defining borders and areas for each group of IFUs. The steps also shows the process of generating a temporary (initial/final) configuration from original (initial/final) configuration.

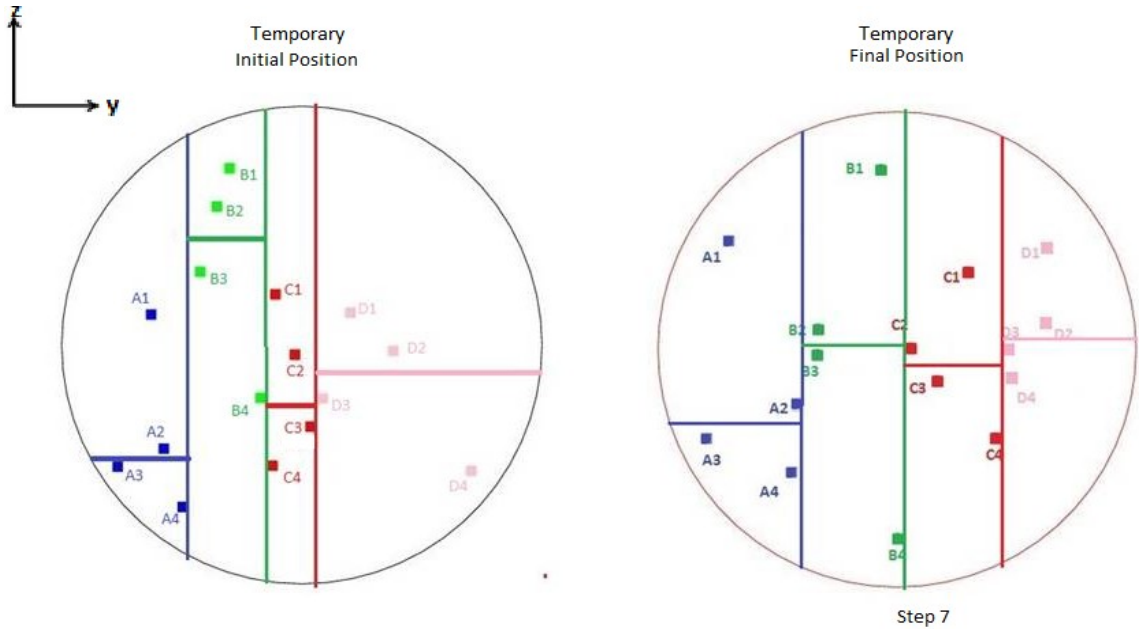

Figure 7: Illustrates movement sequence of sampling elements based on available area. Each sampling element and its movement order is also shown for the target distribution (final configuration) shown in 2.3 and for some dummy initial distribution (current configuration).

- Step 5: For group C, the y coordinate of the highest y SE is defined as the boundary. The boundary is marked with a vertical red line.
- Step 6: Group D SEs D2, D3 and D4 are moved to a temporary position outside the boundary of group C. This forms the temporary final position of all the SEs.
- Step 7: Each group area is bisected into two by drawing a line at the middle of SE 2 and 3 of each group (refer to Figure 7).

In temporary initial and temporary final configurations, each SEs and each group has an associated area (enclosed by the border and/or edge of the field of view). For any SE or any group, this area changes from temporary initial to temporary final configuration. This change in area associated with an SE or a group is important as it defines which SE or group has space to move. Hence, the area change (expand/shrink) defines the sequence of movement. Naturally, an SE or a group whose area shrinks (/shrinks more/ expands less) from temporary initial to temporary final configuration should move before the one whose area expands (/shrinks less/expands more). A typical motion sequence generated by the algorithm between two mock distributions is shown in Figure 7. The following steps determines the sequence of movement of SEs from temporary initial position to temporary final positions for the Figure 7. Here shrink or expand means shrink or expand from temporary initial configuration to temporary final configuration.

- Step 1: Combined area of C and D shrinks while that of A and B expands. So SEs of C and D should move first. Out of 16 movement sequence 1-8 is allocated to C and D while 9-16 is allocated to A and B where 1 and 16 defines the first and the last to move respectively.
- Step 2: Between C and D, the area of D shrinks while that of C expands. So the D SEs would have sequence 1-4 and the C SEs will have the sequence 5-8.
- Step 3: Within D SEs, the area of D1 and D2 Shrinks. So D1 and D2 will have sequence 1-2 while D3 and D4 will have 3-4.
- Step 4: In the same manner D1 should have sequence 1 and D2 should have sequence 2. While D3 and D4 should have sequence 3 and 4 respectively.
- Step 5: Continuing this process we can determine the movement sequence of all 16 SEs.

In the final step of the algorithm, the movement sequence is handed over to the control system firmware in a text format. The primary job of the firmware is to store the current positions of all the SEs and to move the linear stages. It also sends these current positions at a frequent interval to the algorithm software, to ensure that the $S^I$ is updated. After receiving the movement sequence, the firmware starts moving the OAs in a quasi-parallel manner (two or four at a time), reducing the time required for deployment. In quasi parallel motion, up to four SEs of a group can move together. The algorithm determines whether a quasi parallel motion is allowed if the z coordinates of the initial and final position of SEs within a group does not cross. In the above example movement of group C SEs are eligible for quasi parallel movement. Sequential and parallel movements are illustrated in Figure 8 and 9 respectively.

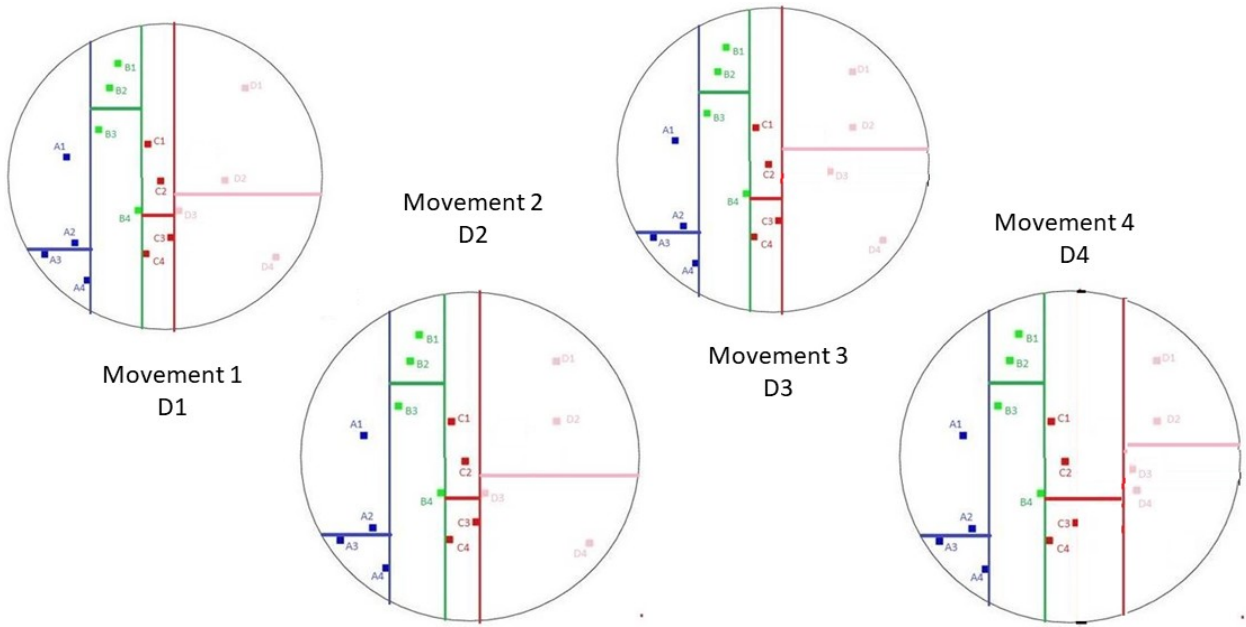

Figure 8: Illustrates shows sequential movement of group D SEs f the scenario depicted in Figure 7.

The Geometry imposes some constraints in terms of the maximum number of objects with overlapping z coordinates. Rotating the field in small steps by turning the Cassegrain port of the telescope overcomes this difficulty.

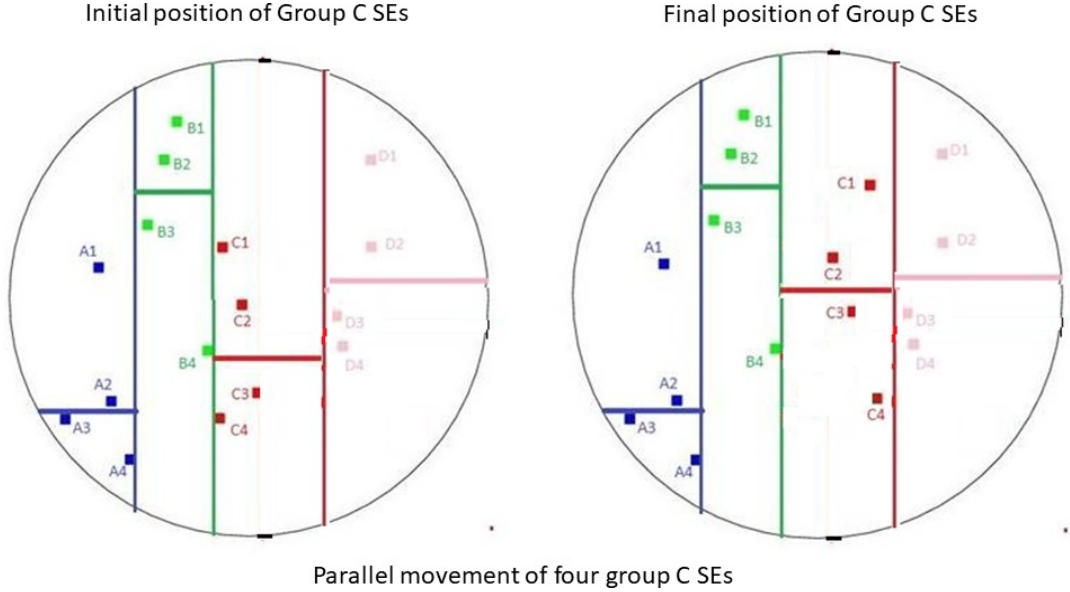

Figure 9: Illustrates shows parallel movement of group C SEs of the scenario depicted in Figure 7.

# 3 SIMULATION RESULTS

The algorithm has been implemented in Python, and it consists of three segregated blocks of code. The first block takes inputs from the user via a file or a database about the target positions in the sky, and the available SEs. It then converts these object sky coordinates (in RA/DEC) into mechanical coordinates (in z/y mm). The second code block makes the assignment and movement sequence. The third block checks the movement sequence and reports on any possibility of a collision during the motion. Once the sequence is verified to be collision free, the movement sequence is sent to the firmware. In all the blocks, a preliminary check of target accessibility is implemented before the application of the algorithm.

We have used a python tool to generate random distributions of objects across the field to verify the performance of the MRC scheme. The simulation study is undertaken to serve the need of DOTIFS but can be extended beyond that scope. Some typical target fields are shown in Figure 10 for values of $\tau$ (the ratio of number of targets to number of available SEs) varying from 0.25 to 6.

For each value of $\tau$, we have created a million target distributions where the targets are distributed across the field of view. Then the algorithm is used to devise an assignment and a movement sequence of IFUs between two mock distributions. For any distribution, the algorithm generates a set of intermediate positions (frame) of the SEs every 5 seconds. This process takes <2 s for 16 SEs and 5 s for 150 SEs.?? The timings of these frames are defined by the travel speed of the OAs. For 16 SEs, the algorithm was successful in every trial of the one million mock distributions. It is found that, about 23% of the time, the mechanical constraint of the Geometry created the stumbling block. However, failure in acquisition was in each case solved by rotating the field by steps of 2° and by searching for an optimal solution in the rotated frame. Acquisition of any object in the FOV without the need for prioritization of the objects is a major advantage of the MRC. Also, since the SEs can move simultaneously the reconfiguration time is considerably reduced.

MRC is designed to perform equally well for both distributed and clustered (extended) targets. Similar to distributed targets, we have performed the simulation for clustered targets. A configuration with clustered targets will have at least 5 or more targets adjacent to each other. We have crated one million distributions with at least one cluster of targets. The GAC has successfully allocated the SEs to the objects (both clustered and sparse) and moved them by avoiding any collision for all scenarios. Thus, the allocation of SEs to a clustered object field does not pose any a limitation to the system.

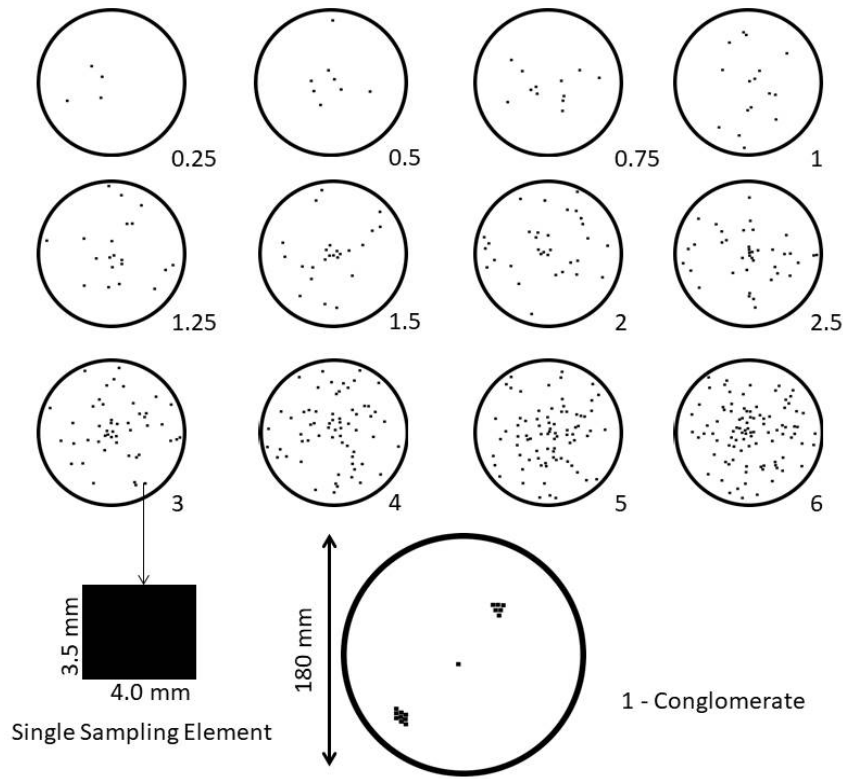

Figure 10: Simulated distribution of objects over the field of view. The numbers denote the value of $\tau$ for that distribution for 16 SEs. Lower left figure shows the dimension of DOTIFS SE and the focal plane with a clustered target distribution.

The GAC can be implemented for a higher number of SEs and a larger dimension of the focal plane. Figure 11 shows the effect of the number of SEs on maximum reconfiguration time for different sizes of the FOV. It is assumed that the average speed of an linear stage is 20mm/s which is achievable by using a stepper motor driven or a piezo stage. The deployment time increases linearly with the number of SEs as shown in figure 11. Measured from the simulation, on average the MRC requires ~130 (/25) seconds on average to deploy 100 (/16) SEs on a focal plane of 200 (/20) cm. MRC is faster than most of the present day methodologies like Starbugs which takes 5 minutes to reconfigure a field of diameter 327 mm for 150 SEs. For comparison, MRC takes 120 seconds to deploy 150 SEs in a field of 350 mm diameter. This advantage can be primarily attributed to speed of linear positioners.

## 3.1 Advantages

The MRC method provides several advantages over the existing GACs. These advantages that are relevant to all types of astronomical observations, not limited to only spectroscopy.

### 3.1.1 Unbiased Sky Coverage

The issue has already been discussed for the fishermen round pond Geometry in the section 1. This necessitates prioritization of the objects. The MRC method provides the freedom to observe any object within the field without the need for prioritization. In the cases where $\tau$ is greater than unity, the prioritization of objects is also acceptable but not necessary.

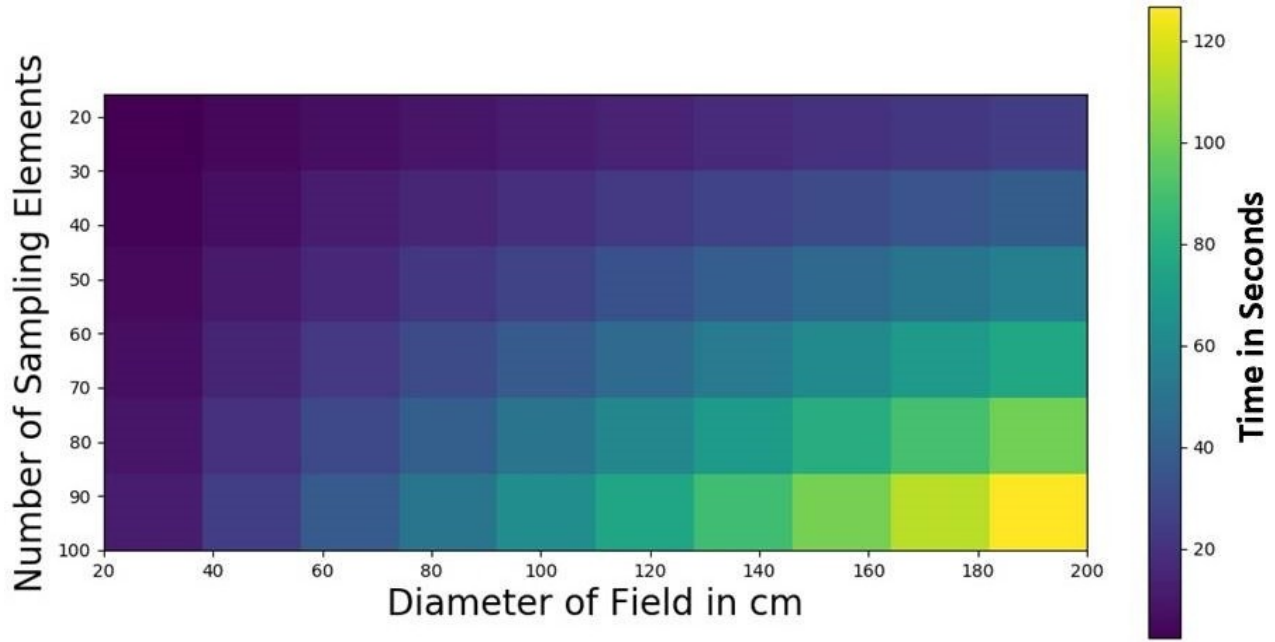

Figure 11: Average time required for reconfiguration with varying number of Sampling Elements and field diameter for $\tau = 1$.

### 3.1.2 Field Revisit Efficiency

The field revisit efficiency of any GAC has a strong dependence on the object density of a field. Both sequential and parallel GACs suffer the problem of inefficient target acquisition for a larger number of revisits of the same field. A parallel scheme of discretely controlled fibers like Starbugs and MRC method can not be directly compared as Starbugs will have more number of SEs but each SE has way less number of spaxels compared to the MRC. Hence the Starbug test target distribution needs to be structured into targets of size that can be covered by multi spaxel IFU. The efficiency of the MRC on this restructured field can be compared to the Starbugs efficiency. Figure 12 shows a comparison of the efficiency in a field revisit between the MRC and the Starbugs. The trend shows the efficiency of 151 Starbug robots is dropping considerably after the eighth visit of the field (shown in Figure 13) containing 1431 targets. A similar target distribution is created with same number of targets within a field of view of diameter 327 mm which matches that of TAIPAN focal plane. The MRC using 144 SEs is found to acquire all the targets within the 10th visit. This shows that visiting the target area multiple times does not impart any effect on the efficiency of the MRC as compared to the Starbugs GAC. MRC benefits the advantage of it's unique geometry which makes it more efficient for field revisit and attending clustered targets. On the other hand, the geometry is the primary constraint here for the introduction of inefficiency in case of Starbugs. A similar effect can be seen for other parallel techniques as well [5].

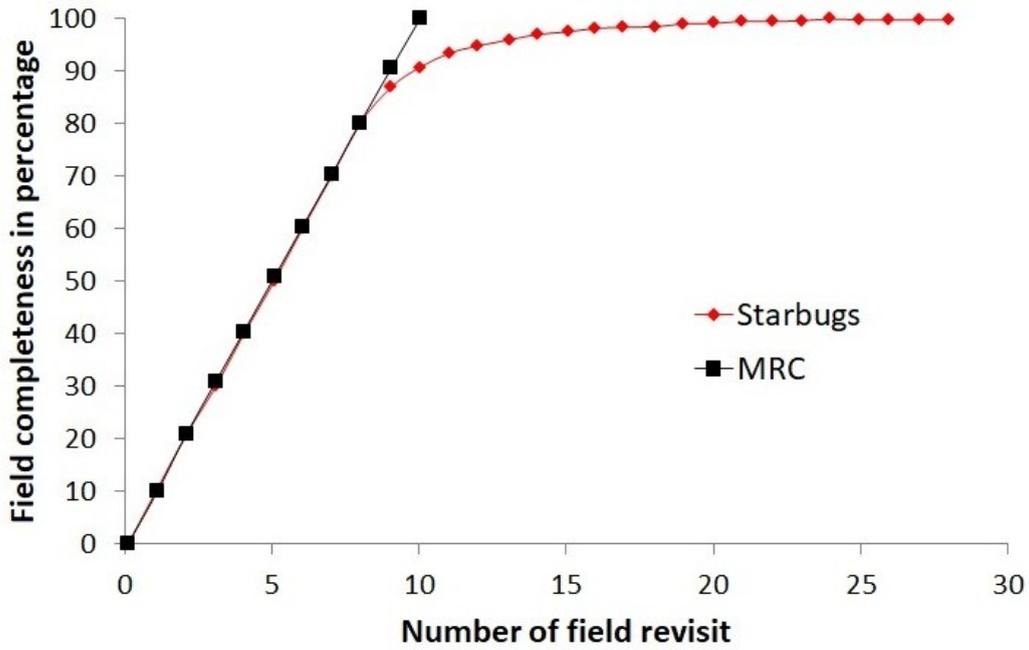

Figure 12: Completeness of target acquisition against number of field visits shows that the efficiency of Multiple Rooks of Chess does not change with multiple revisit. The simulation is performed for MRC with 1431 targets nd 144 sampling elements for a 327mm diameter circular field of view.

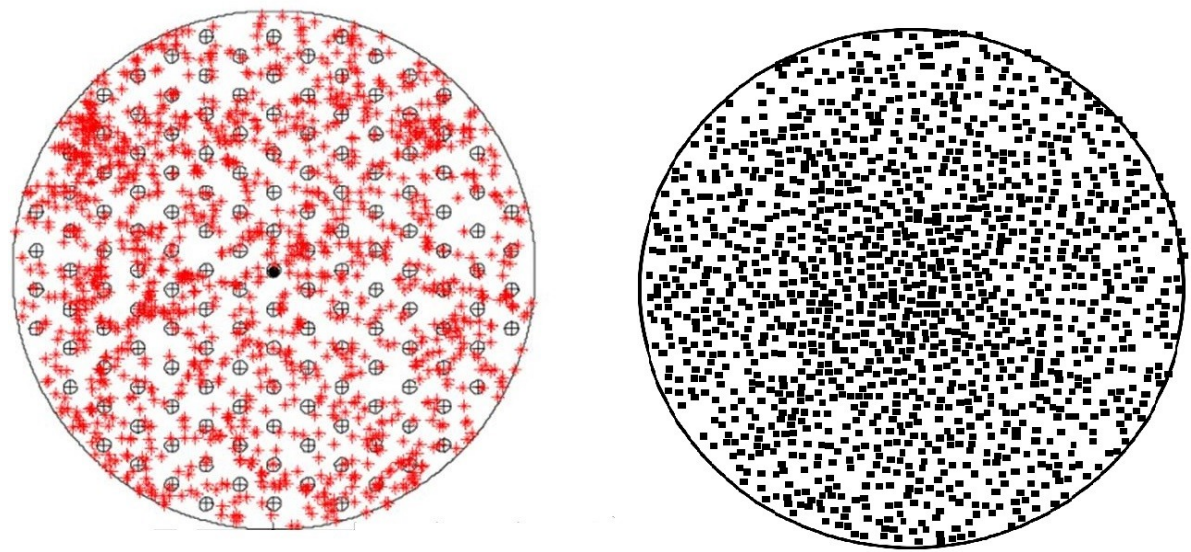

Figure 13: Left: Target field configuration of 1431 objects (marked as red star) and 151 Starbug robots (marked as black hollow circle) over a field of 327mm diameter. The same field is restructured into 157 objects on which MRC is deployed to compare the efficiency. The image is taken from [6]. Right: Target distribution with same number of targets which is used to characterize the efficiency of MRC method. Each box represents a target area which can be covered by an IFU.

### 3.1.3 Contiguous Target Acquisition

Several of the science cases of the DOTIFS are based on the observation of extended objects by contiguously sampling them. In a GAC with densely populated SEs, it is possible to sample an extended object uniformly using slightly dithering the telescope. However, the process of multiple observation via telescope dithering is very inefficient which is a major drawback. For the fastest fiber deployment techniques like the bend-arm or the spine, the minimum separation between any two SEs is typically multiple arcseconds (12" and 100" for Echidna and Starbugs respectively). In our GAC, although an SE may or may not have multiple fibers, they can be placed contiguously within a space of 0.5" in the sky. The separation between two fibers within the SE is less than the inter SE gap for a multi-fiber SE. Table 1 shows the comparison of the MRC method with different techniques and an upgraded version of the GAC based on the number of sampling elements as well as the field diameter. It can easily be inferred that the MRC method provides the best contiguous target acquisition.

| Parameters | Multiple Rooks of Chess for DOTIFS | Multiple Rooks of Chess Upgrade | Starbugs for TAIPAN | Starbugs for MANIFEST | Bend arm robots MOONS |
|---|---|---|---|---|---|
| Number of Sampling Elements | 16 | 100 | 150 | 600 | 1000 |
| Field Diameter in mm | 180 | 2000 | 327 | 1250 | 880 |
| Speed of positioners in mm/sec | 20 | 20 | 2 | 2 | - |
| Reconfiguration Time in seconds | ~25 | ~127 | <300 | <480 | ~300 |
| Minimum edge-to-edge distance between sampling elements in mm | 0.1 | 0.1 | 9 | 9 | ~17 |
| Patrol area of sampling elements in sq cm | ~250 | ~$3 \times 10^4$ | ~80 | ~300 | ~7 |

Table 1: Comparison between the different geometry and algorithm combinations in terms of field reconfiguration parameters. Multiple Rooks of Chess Upgrade column shows the performance of an upgraded version of the geometry and algorithm combination in terms of the number of sampling elements as well as field diameter. Details of TAIPAN, MANIFEST and MOONS are derived from [11], [12] and [8] respectively.

### 3.1.4 Deployment Overhead Time

In modern day large telescopes, the cost of observation time is enormous. Some of the sequential GACs take close to an hour (Simulated Annealing in 2dF survey) to reconfigure the field. Although they use multiple plates and positioners and minimize the time of reconfiguration by configuring one focal plane plate while the other plate used for observation, their scheme is inefficient in terms of cost. At this point, Starbugs (which is the fastest parallel technique) requires 300 seconds to assign 150 fibers on a field of 327 mm diameter on the TAIPAN focal plane. As mentioned before, MRC can reconfigure a field of 350 mm diameter with 150 SEs in

~120 seconds on average. The deployment time changes only linearly if the size of the field or the number of SEs changes, as shown in table 1.

## 4 SUMMARY

There are several methods of field reconfiguration which are sequential and parallel in nature. These methods has different inherent limitations such as central target bias, inefficiency in covering clustered targets etc. The limitations arise due to their mechanical configuration. The associated algorithms are complex as they have to prioritize the targets to avoid these limitations. We propose a GAC that is free from these inherent limitations as well as generic enough to be suitable for a survey or a specific observation.

Multiple Rooks of Chess as GAC is demonstrated to be extremely flexible in attaining any target configuration of different $\tau$ value. For a 2m focal plane (typical for 30m class telescope) with 100 SEs, the reconfiguration time can be maximum up to 127 seconds including algorithm and deployment. Hence it is much faster than all the currently available fiber deployment schemes. It can successfully attain all the targets of any target distribution without the requirement to prioritize the objects. This is independent of the number of fibers in a single Sampling Element. One constraint in the form of the limitation of the number of identical y-coordinate targets can be avoided by introducing field rotation, which is not an issue for integral field spectroscopy.